\documentclass[prd, preprint, longbibliography, 12pt]{revtex4-1}

\usepackage{amsmath}
\usepackage{amssymb}
\usepackage{setspace}
\usepackage{graphicx}
\usepackage{natbib}
\usepackage{float}
\usepackage{tensor}
\usepackage[utf8]{inputenc}
\usepackage{amsfonts}
\usepackage{braket}
\usepackage{esint}
\usepackage{breqn}
\usepackage{IEEEtrantools}

\begin{document}
 
 %

\begin{center}
{ \large \bf Proposal for a new quantum theory of gravity III\\
\large \it - Equations for quantum gravity, and the  origin of spontaneous localisation -\\
 }


\vskip 0.2 in

{\large{\bf Palemkota Maithresh$^{a1}$ and Tejinder P.  Singh$^{b2}$ }}

\medskip

{\it $^a${UM-DAE Centre for Excellence in Basic Sciences, Mumbai, 400098, India}\\

{\it $^b$Tata Institute of Fundamental Research,}
{\it Homi Bhabha Road, Mumbai 400005, India}\\
\bigskip
 {$^{1}$\tt p.maithresh@cbs.ac.in}, \; {$^2$\tt tpsingh@tifr.res.in}}

\end{center}

\centerline{\bf ABSTRACT}
\noindent  We present a new, falsifiable, quantum theory of gravity, which we name Non-commutative Matter-Gravity. The commutative limit of the theory is classical general relativity.
In the first two papers of this series, we have introduced the concept of an atom of space-time-matter [STM], which is described by the spectral action in non-commutative geometry,  corresponding to a classical theory of gravity. We used the Connes time parameter, along with the spectral action, to incorporate gravity into trace dynamics. We then derived the spectral equation of motion for the gravity part of the STM atom, which turns out to be the Dirac equation on a non-commutative space. In the present work, we propose how to include the matter (fermionic) part and give a simple action principle for the STM atom. This  leads to the equations for a quantum theory of gravity, and also to an explanation for the origin of spontaneous localisation from quantum gravity. We use spontaneous localisation to arrive at the action for classical general relativity [including matter sources] from the action for STM atoms.

\bigskip

This paper should ideally be read as a follow-up to the first two papers in this series \cite{Singh2019qg, Singh2019qgii}, which will be hereafter referred to as I and II respectively.

In I, we have introduced the concept of an atom of space-time-matter [STM], which is described by the spectral action of non-commutative geometry. The spectral action,  in the presence of a Riemannian manifold, is equal to the Einstein-Hilbert action of classical general relativity, after a heat kernel expansion of square of Dirac operator is carried out, and truncated at the second order in an expansion in $L_p^{-2}$. We also introduced there the four levels of gravitational dynamics.
In II, we used the Connes time parameter, along with the spectral action, to incorporate gravity into trace dynamics. We then derived the spectral equation of motion for the gravity part of the STM atom, which turns out to be the Dirac equation on a non-commutative space.  In the present work, we propose how to include the matter (fermionic) part and give a simple action principle for the STM atom. This  leads to the equations for a quantum theory of gravity, and also to an explanation for the origin of spontaneous localisation from quantum gravity. We use spontaneous localisation to arrive at the action for classical general relativity [including matter sources] from the action for STM atoms.

\section{The equations of quantum gravity, at Level 0}
In II, we have proposed the following action principle for the gravity part of the STM atom:
\begin{equation}
S_{GTD} =\kappa \frac{c}{L_P} \int d\tau \; Tr [\chi (L_P^2 \; \dot{q}^2/L^2 c^2)]
\label{acgtd}
\end{equation}
Here, $\tau$ is what we have called Connes time of non-commutative geometry. The $q$ operator which describes gravity, is related to the operator $D$ (which becomes the standard Dirac operator  on a curved space when there is a background Riemannian manifold) as follows:
\begin{equation}
D \equiv \frac{1}{Lc}\;  \frac{dq}{d\tau}
\label{dirq}
\end{equation}
The function $\chi(u)$ is so chosen as to ensure convergence of the heat kernel expansion of $Tr(L_p^2 D^2)$ [for a discussion on this aspect, see e.g. \cite{Connes2000}, \cite{Landi1999}]. $\kappa$ is a constant so chosen that it gives the correct dimensions of action, and the correct numerical coefficient for recovery of the Einstein-Hilbert action. $L$ is a length scale associated with the STM atom, whose physical interpretation will become evident subsequently.

Our motivation behind introducing the operator $q$ `particle' is to establish contact between non-commutative geometry [and the description of gravity therein] on one hand, and trace dynamics on the other. We were seeking an action principle which can be expressed conventionally as the time-integral of a Lagrangian, with the Lagrangian being made of matrix-valued configuration variable $q$ and its velocity $\dot q$. Hence the action (\ref{acgtd}), and the relation (\ref{dirq}) which relates the $q$-operator to gravity, via the spectral action and its heat kernel expansion.

The above description, which is the essence of what was done in II, serves as the starting point for the present paper: we will now propose an action principle for the STM atom which includes fermions, in addition to gravity. First, we simplify the above gravity action and its notation. We will assume for now that $\chi(u)=u$, leaving for later the considerations of convergence of the heat kernel expansion. Further, setting $\kappa\equiv C_0 /2$, where $C_0$ is a real constant with dimensions of action, we can write the action (\ref{acgtd}) as
\begin{equation}
\frac{L_P}{c}\frac{S_{GTD}}{C_0} = \frac{1}{2} \int d\tau \; Tr [L_P^2\; \dot q^2/L^2 c^2]
\label{acgtd2}
\end{equation}

$q$ is assumed to have dimension of length, and the expression inside the trace is dimensionless.
In the spirit of trace dynamics, we shall assume that the matrix $q$ (equivalently operator) is made from elements which are complex numbers or anti-commuting Grassmann numbers. In particular, we shall assume that the $q$ matrix above is made from even grade elements of the Grassmann algebra, and is therefore a `bosonic' matrix, which we shall henceforth label as $q_B$. This assumption is natural keeping in view that the above action describes gravity, via the spectral action of non-commutative geometry, and the Dirac operator is bosonic (and self-adjoint). Thus we rewrite the above action (\ref{acgtd}) as
\begin{equation}
\frac{L_P}{c}\frac{S_{GTD}}{C_0} = \frac{1}{2} \int d\tau \; Tr [L_P^2\; \dot q_B^2/L^2 c^2]
\label{acgtd3}
\end{equation}
and relate $q_B$ to the Dirac operator as
\begin{equation}
D_B \equiv \frac{1}{Lc}\;  \frac{dq_B}{d\tau}
\label{dirq2}
\end{equation}

Since the concept of an STM atom was introduced in I as an entity which describes both matter and gravity at Level 0, we must now introduce the fermionic/matter aspect in this action. In order to do so, we define a new $q$-operator as follows:
\begin{equation}
q= q_B + \; q_F
\end{equation}
where $q_F$ is fermionic, i.e. it is made of odd grade elements of the Grassmann algebra.
 However we do not yet place any adjointness requirement on $q_B$ or $q_F$: the Dirac operator $D_B$  will now be made from the self-adjoint part of $q_B$. The above split of $q$ as bosonic plus fermionic simply represents the fact that any matrix made from Grassmann elements can be written as a sum of a bosonic matrix plus a fermionic matrix. The split is significant though, as we will soon see that $q_F$ behaves very differently from $q_B$: not only does it describe emergent fermions, but it also paves the way for spontaneous localisation in a quantum gravity theory. The STM atom is assumed to be described by the following fundamental action principle, which is at the heart of all subsequent development:
 \begin{equation}
\frac{L_P}{c} \frac{S}{C_0}  =  \frac{1}{2} \int d\tau \; Tr \bigg[\frac{L_P^2}{L^2c^2}\; (\dot{q}_B +\beta_1 \dot{q}_F)\;(\dot{q}_B +\beta_2 \dot{q}_F) \bigg]
\end{equation}
Here $\beta_1$ and $\beta_2$ are two constant fermionic matrices. These matrices make the Lagrangian bosonic. The assumptions on these matrices are that they should not both simultaneously commute (or anti-commute) with $\dot{q}_F$ (as justified later in the paper).
These assumptions are necessary for retaining the $\dot{q}_F\dot{q}_F$ term in the trace Lagrangian, which would otherwise vanish. The above trace Lagrangian can be expanded and written as
\begin{equation}
\frac{L_P}{c} \frac{S}{C_0}  =  \frac{a}{2} \int d\tau \; Tr \bigg[ \dot{q}_B^2 +\dot{q}_B\beta_2\dot{q}_F +\beta_1 \dot{q}_F \dot{q}_B +\beta_1 \dot{q}_F \beta_2 \dot{q}_F \bigg]
\label{newacn}
\end{equation}
where we have denoted $a\equiv L_P^2/L^2c^2$.

The first term inside the trace Lagrangian has the familiar structure of a kinetic energy, and in any case is what gives rise to the Einstein-Hilbert action in the heat kernel expansion of $D_B^2$.  It is the cross terms in the trace Lagrangian, which result from introducing the fermionic $q_F$, which are a game changer, and as we shall see,  responsible for causing spontaneous collapse, besides bringing in fermions. As in trace dynamics, we assume the trace Lagrangian to be an even grade element of the Grassmann algebra. We will denote the trace Lagrangian by the symbol ${\bf P}$, i.e. ${\bf P}=Tr\; \cal L$, where $\cal L$ is the operator polynomial which defines the Lagrangian in this matrix dynamics.

It is noteworthy that the introduction of the two constant matrices $\beta_1$ and $\beta_2$ seems essential, for the following reasons. Our starting point for constructing the present Lagrangian is the gravity Lagrangian in (\ref{acgtd2}) for the bosonic $q_B$. It is natural that to introduce fermions, we generalise $q_B$ to $q=q_B + q_F$. We also expect the Lagrangian to be quadratic in time derivative with respect to $\tau$, and we also ask for the Lagrangian to be bosonic. This makes it essential that a constant fermionic matrix $\beta$ be brought in, and the trace Lagrangian be made from the bosonic $q=q_B + \beta q_F$. For instance, the trace Lagrangian could be $Tr (\dot q_B + \beta \dot q_F)^2$. Intriguingly, we did not succeed in making a consistent model with only one constant matrix in the Lagrangian. On the other hand, the situation eases immediately when two constant matrices are brought in (i.e. $\beta_1$ and $\beta_2$). Furthermore, if we are seeking a bosonic trace Lagrangian which is at the most quadratic in $q$, then it will not be possible to introduce more than two constant matrices - so we seem to be dealing with a generic `free-particle' quadratic trace Lagrangian, which incorporates $q_F$. Our Lagrangian is not self-adjoint (nor the action is), though as we shall see, it becomes self-adjoint in the limit in which classical dynamics [Level III] and quantum field theory [Level II] are recovered. The anti-self-adjoint part of the Lagrangian is responsible for spontaneous localisation, and it arises quite naturally from the structure of our assumed trace Lagrangian: soon as the fermionic part $q_F$ is introduced, spontaneous localisation becomes inevitable.

There are three universal constants in the theory: Planck length $L_P$ and Planck time $\tau_P=L_P/c$, where the speed of light $c$ should be thought as the ratio $L_p/\tau_P$. The third universal constant $C$ has dimensions of action, and at Level I will be identified with Planck's constant $\hbar$.
Newton's gravitational constant $G$ and Planck mass $m_P$ are emergent only at Level I. In fact the concepts of mass and spin themselves emerge only at Level I, and are not present at Level 0. We  associate only a length scale (more precisely an area $L^2$) with the STM atom, but not mass nor spin, at Level 0.

One can now derive the Lagrange equations of motion, as is done in trace dynamics. The derivative of the trace Lagrangian ${\bf P}$ (note that ${\bf P}$ is a complex number) with respect to an operator $\cal O$ in ${\cal L}$ is defined as
\begin{equation}
\delta {\bf P }= Tr \frac{\delta{\bf P}}{\delta\cal{O}}\delta\cal{O}
\end{equation}
This so-called trace derivative is obtained by varying ${\bf P}$ with respect to ${\cal O}$ and then cyclically permuting ${\cal O}$ inside the trace, so that $\delta\cal O$ sits to the right of the polynomial $\cal{L}$. While permuting cyclically inside the trace, one has to keep in mind the change in sign when permuting two fermionic matrices $\chi_1, \chi_2$, and no change in sign when a bosonic matrix $B$  is permuted with any other matrix:
\begin{equation}
Tr [B_1, B_2] = Tr [B_2, B_1], \qquad Tr [B, \chi] = Tr [\chi, B], \qquad Tr [\chi_1, \chi_2] = - Tr [\chi_2\chi_1] 
\label{adjr}
\end{equation}

The extra sign that appears in the commutator of fermionic matrices in Eqn. (\ref{adjr})  causes these matrices to follow different adjointness properties. If $\mathcal{O}_1^{g_1}$,...,$\mathcal{O}_{n}^{g_n}$ are $n$ matrices with grades $g_1$,..,$g_n$ respectively, then  
\begin{equation}
    (\mathcal{O}_1^{g_1}...\mathcal{O}_n^{g_n})^\dagger = (-1)^{\sum_{i<j}g_ig_j} {\mathcal{O}_n^{g_n}}^\dagger ... {\mathcal{O}_1^{g_1}}^\dagger 
\end{equation}
So, two fermionic matrices $\chi_1$ and $\chi_2$ obey $(\chi_1 \chi_2)^\dagger = -\chi_2^\dagger \chi_1^\dagger$. This minus sign is not there  if one or both matrices are bosonic. 

We can now vary the action (\ref{newacn}) with respect to $q_B$ and $q_F$, in the spirit of trace dynamics,  and obtain the Lagrange equations of motion:
\begin{equation}
\frac {d\;}{d\tau}\Big ( \frac{\delta {\bf P}}{\delta \dot q_B}\Big ) - \Big ( \frac {\delta {\bf P}}{\delta q_B}\Big ) = 0
\end{equation} 
and an analogous equation for $q_F$. Since the trace Lagrangian  is independent of $q$, the conjugate momenta $p_B=  {\delta {\bf P}}/{\delta \dot q_B}$ and  $p_F=  {\delta {\bf P}}/{\delta \dot q_F}$ are constant. From the trace derivative of the trace Lagrangian with respect to $\dot{q}_B$ and $\dot{q}_F$ we get the momenta to be 
\begin{align}
    p_B = \frac{\delta \textbf{L}}{\delta \dot{q}_B} &= \frac{a}{2}\bigg[2\dot{q}_B +(\beta_1 +\beta_2)\dot{q}_F \bigg]\\ 
    p_F = \frac{\delta \textbf{L}}{\delta \dot{q}_F} &= \frac{a}{2}\bigg[\dot{q}_B (\beta_1 +\beta_2)+\beta_1 \dot{q}_F \beta_2 + \beta_2 \dot{q}_F \beta_1 \bigg]
\end{align}
We note that all the degrees of freedom $q_B, q_F, p_B, p_F$ obey arbitrary time-dependent commutation relations with each other. Quantum commutation relations emerge after constructing  a statistical thermodynamics for an ensemble of STM atoms \cite{Adler:04}. 

 The momenta $p_B$ and $p_F$ are respectively bosonic / fermionic.
Both the momenta are constant, because the trace Lagrangian does not depend on $q$. This implies, 
\begin{align}
    2\dot{q}_B +(\beta_1+\beta_2)\dot{q}_F = c_1 \\
    \dot{q}_B(\beta_1+\beta_2)+ \beta_1 \dot{q}_F \beta_2 +\beta_2 \dot{q}_F \beta_1 = c_2
\end{align}
where $c_1$ and $c_2$ are constant bosonic and fermionic matrices, respectively.
These equations yield the following solutions for $q_B$ and $q_F$:
\begin{align}
     \dot{q}_B &= \frac{1}{2}\bigg[c_1 -(\beta_1+\beta_2)(\beta_1-\beta_2)^{-1}\big[2c_2 -c_1(\beta_1+\beta_2) \big](\beta_2-\beta_1)^{-1} \bigg] \label{qb} \\
     \dot{q}_F &= (\beta_1-\beta_2)^{-1}\big[2c_2-c_1(\beta_1+\beta_2)\big](\beta_2-\beta_1)^{-1} \label{qf}
\end{align}
This means that the velocities $\dot{q}_B$ and $\dot{q}_F$ are constant,  and $q_B$ and $q_F$ evolve linearly in Connes time. 

Since $p_B =\frac{a}{2}c_1$ and $p_F = \frac{a}{2}c_2$, [\ref{qb}] and [\ref{qf}] can be written as 
\begin{align}
    \dot{q}_B &= \frac{1}{a}\bigg[p_B -(\beta_1+\beta_2)(\beta_1-\beta_2)^{-1}\big[2p_
    F-p_B(\beta_1+\beta_2) \big](\beta_2-\beta_1)^{-1} \bigg] \\
     \dot{q}_F &= \frac{2}{a}(\beta_1-\beta_2)^{-1}\big[2p_F-p_B(\beta_1+\beta_2)\big](\beta_2-\beta_1)^{-1}
\end{align}

The trace Hamiltonian ${\bf H}$ can be constructed as
\begin{equation}
{\bf H}=\mathrm{Tr}[p_{F}\dot{q}_{F}]+\mathrm{Tr}[p_{B}\dot{q}_{B}]-\mathrm{Tr}\; \mathcal{L}
\end{equation}
which becomes, after substituting for momenta and the Lagrangian,
\begin{equation}
   \textbf{H} = \text{Tr}\bigg[\frac{a}{2} (\dot{q}_B+\beta_1\dot{q}_F)(\dot{q}_B+\beta_2\dot{q}_F) \bigg]
\end{equation}
and in terms of the momenta
\begin{equation}
    \textbf{H} = \text{Tr} \frac{2}{a} \bigg[(p_B\beta_1-p_F)(\beta_2-\beta_1)^{-1}(p_B\beta_2-p_F)(\beta_1-\beta_2)^{-1}
    \bigg]
\end{equation}

In trace dynamics,  Hamilton's equations of motion are 
\begin{align}
    \frac{\delta \textbf{H}}{\delta q_r} =-\dot{p}_r, \quad \quad \frac{\delta \textbf{H}}{\delta p_r} =\epsilon_r \dot{q}_r
\end{align}
where $\epsilon_r = 1(-1)$ when $q_r$ is bosonic(fermionic). 
For our case, the Hamilton's equations for bosonic variables are
\begin{IEEEeqnarray}{rCll}
\dot{q}_B &=& \frac{2}{a} \Big[&\beta_1 (\beta_2-\beta_1)^{-1} (p_B\beta_2 -p_F)(\beta_1-\beta_2)^{-1} \nonumber\\
    &&&+\beta_2 (\beta_1-\beta_2)^{-1}(p_B\beta_1-p_F)(\beta_2-\beta_1)^{-1} \Big] \\
    \dot{p}_B &=& 0&
\end{IEEEeqnarray}
The Hamilton's equations for fermionic variables are
\begin{IEEEeqnarray}{rCll}
\dot{q}_F &=& -\frac{2}{a}\Big[&(\beta_2-\beta_1)^{-1}(p_B\beta_2-p_F)(\beta_1-\beta_2)^{-1}   \nonumber \\
    &&&+(\beta_1-\beta_2)^{-1}(p_B\beta_1-p_F)(\beta_2-\beta_1)^{-1} \Big] \\
     \dot{p}_F &=&0&
\end{IEEEeqnarray}
It can be verified that these equations are identical with those solutions above which come from Lagrange's equations.

Taking cue from the expression for $p_B$ we can define the generalised (bosonic) Dirac operator $D$ given by
\begin{equation}
\frac{1}{Lc}\;  \frac{dq}{d\tau}\sim D \equiv D_B + D_F ; \qquad D_B \equiv \frac{1}{Lc}\;  \frac{dq_B}{d\tau} ; \qquad   D_F \equiv \frac{\beta_1 + \beta_2}{2Lc} \frac{dq_F}{d\tau}
\end{equation} 
We note that it is a constant operator, and we can also express this as an eigenvalue equation
\begin{equation}
[D_B + D_F] \psi = \lambda \psi
\end{equation}
where the eigenvalues $\lambda$, assumed to be $c$-numbers, are independent of Connes time $\tau$, and the state $\psi$ can depend on $\tau$ at most through a multiplicative factor.

In trace dynamics, there is a conserved charge, known as the Adler-Millard charge, corresponding to a global unitary invariance of the trace Lagrangian / Hamiltonian. Assume a dynamical operator $x_{r}$ undergoes a  transformation as $x_{r}\rightarrow U^{\dagger}x_{r}U$
where $U$ is a constant $N\times N$ matrix, given by
 $U=\exp\Lambda$, where $\Lambda$ is an anti-self-adjoint bosonic generator matrix. Under such a transformation of operators, the trace Hamiltonian remains invariant.

Thus, the Adler-Millard charge is conserved  under the transformations which obey
\begin{equation}
    \label{gui}
    \textbf{\cal L}(\{U^\dagger q_r U\}, \{U^\dagger \dot{q}_r U\}) = \textbf{\cal L}(\{q_r\},\{\dot{q}_r\})
\end{equation}
where $U$ is a constant unitary matrix, which is written as $U=\exp{\Lambda}$, where $\Lambda$ is an anti-self-adjoint bosonic generator matrix.  Applying the above condition on our Lagrangian gives 
\begin{multline}
    \textbf{L}(\{U^\dagger q_r U\}, \{U^\dagger \dot{q}_r U\}) = \text{Tr}\bigg[U^\dagger \dot{q}_B^2 U +U^\dagger \dot{q}_B U \beta_2 U^\dagger \dot{q}_FU \\
    +\beta_1U^\dagger \dot{q}_F \dot{q}_B U +\beta_1 U^\dagger \dot{q}_F U \beta_2U^\dagger \dot{q}_F U\bigg]
\end{multline}
     The above equation satisfies [\ref{gui}]  if we choose
\begin{equation}
    U\beta_2U^\dagger = \beta_2 \quad \& \quad U\beta_1U^\dagger = \beta_1
\end{equation}
This condition also means that $\beta_1$ and $\beta_2$ commute with U (or $\Lambda$ equivalently)

The Adler-Millard charge in trace dynamics can be shown to be \cite{Adler:04}
\begin{equation}
    \tilde{C} = \sum_{r\in B}[q_r,p_r] -\sum_{r\in F} \{q_r,p_r \} 
    \label{amc}
\end{equation}
Substituting the momenta in the above equation, we get 
\begin{align}
     (2/a) \; \tilde{C} &= [q_B, 2\dot{q}_B +(\beta_1+\beta_2)\dot{q}_F] -\{q_F, \dot{q}_B(\beta_1 +\beta_2)+\beta_1 \dot{q}_F \beta_2+ \beta_2 \dot{q}_F \beta_1 \} \nonumber\\
      &= [q_B,2\dot{q}_B]+[q_B,(\beta_1+\beta_2)\dot{q}_F]-\{q_F,\dot{q}_B(\beta_1+\beta_2)\} \\ \nonumber &\quad \: -\{q_F,\beta_1\dot{q}_F\beta_2+\beta_2\dot{q}_F\beta_1 \} 
\end{align}
  The cross terms in the charge are expected to vanish at equilibrium when one constructs a statistical thermodynamics for this matrix dynamics. The important terms that lead to emergence of statistical thermodynamics, and cause spontaneous collapse, are $ 2[q_B,\dot{q}_B]$ and  $\{q_F, \beta_1 \dot{q}_F \beta_2+\beta_2 \dot{q}_F \beta_1 \}$. 
Splitting $q_B$ into its self-adjoint and anti-self-adjoint parts, i.e. $q_B =q_{BS}+q_{BAS}$, we get 
\begin{equation}
    2[q_B,\dot{q}_B] = 2\bigg([q_{BS},\dot{q}_{BS}]+[q_{BS},\dot{q}_{BAS}]+[q_{BAS},\dot{q}_{BS}]+[q_{BAS},\dot{q}_{BAS}] \bigg)
\end{equation}
The terms $[q_{BS},\dot{q}_{BAS}]$ and $[q_{BAS},\dot{q}_{BS}]$ are self-adjoint and the terms $[q_{BS},\dot{q}_{BS}]$, $[q_{BAS},\dot{q}_{BAS}]$ are anti-self-adjoint. 
Now writing $p_F^{f} = \beta_1 \dot{q}_F\beta_2 +\beta_2 \dot{q}_F \beta_1 $ and splitting $q_F$ and $p_F^f$ into their self-adjoint and anti-self-adjoint parts, i.e. $q_F = q_{FS}+q_{FAS}$ and $p_F =p_{FS}^f +p_{FAS}^f$, we have 
\begin{equation}
 \{q_F,p_F^f \} = \{q_{FS},p_{FS}^f\}+\{q_{FS},p_{FAS}^f\}+\{q_{FAS}, p_{FS}^f\}+\{q_{FAS},p_{FAS}^f\}
\end{equation}
The terms $\{q_{FS},p_{FAS}^f\}$, $\{q_{FAS}, p_{FS}^f\}$ are self-adjoint and the terms \{$q_{FS},p_{FS}^f\}$, $\{q_{FAS},p_{FAS}^f\}$ are anti-self-adjoint. The anti-self-adjoint part of $\tilde{C}$ determines the emergent quantum commutators at equilibrium \cite{Adler:04}.

\subsection{Adjointness Properties}
The Hamiltonian is
\begin{equation}
   \textbf{H} = \text{Tr}\bigg[\frac{a}{2} (\dot{q}_B+\beta_1\dot{q}_F)(\dot{q}_B+\beta_2\dot{q}_F) \bigg]
\end{equation}

It is very important to retain $\text{
Tr}[\beta_1 \dot{q}_F \beta_2 \dot{q}_F]$ in the trace Hamiltonian to get fermionic anti-commutator in the conserved charge:
\begin{align}
    \text{Tr}[\beta_1 \dot{q}_F \beta_2 \dot{q}_F] =& \text{Tr}[-\dot{q}_F\beta_1 \dot{q}_F \beta_2 ] \\
    =& -Tr\Big[\beta_1 \dot{q}_F \beta_2 \dot{q}_F+[\dot{q}_F,\beta_1]\beta_2 \dot{q}_F  \nonumber \\ &+\beta_1\dot{q}_F[\dot{q}_F,\beta_2]+ [\dot{q}_F,\beta_1][\dot{q}_F,\beta_2] \Big] \label{trace1} \\
    =& -\text{Tr}\Big[ \beta_1 \dot{q}_F \beta_2 \dot{q}_F -\beta_1 \dot{q}_F \{\dot{q}_F,\beta_2\}-\{\dot{q}_F,\beta_1 \} \beta_2 \dot{q}_f \nonumber \\
    &+\{\dot{q}_F,\beta_1\}\{\dot{q}_F,\beta_2\}\Big] \label{trace2}
\end{align}
From [\ref{trace1}], [\ref{trace2}], to retain the $\text{
Tr}[\beta_1 \dot{q}_F \beta_2 \dot{q}_F]$, both $\beta_1$ and $\beta_2$ can not simultaneously commute or anti-commute with $\dot{q}_F$. 

The Hamiltonian can be split into its self-adjoint and anti-self-adjoint parts as follows:
\begin{align}
     \textbf{H}_S &= \text{Tr}\Big[\dot{q}_B^2 +
   \big[ (\beta_1\dot{q}_F)_S +(\beta_2\dot{q}_F)_{S}\big]\dot{q}_B+(\beta_1\dot{q}_F \beta_2 \dot{q}_F)_S\Big] \\
   \textbf{H}_{AS}&= \text{Tr}\Big[  \big[(\beta_1 \dot{q}_F)_{AS}+(\beta_2 \dot{q}_F)_{AS} \big]\dot{q}_B +(\beta_1\dot{q}_F \beta_2\dot{q}_F)_{AS}\Big]
\end{align}

Let us consider the adjointness property of the momentum $p_B$. For the remainder of our analysis, we shall assume that $q_B$ is self-adjoint - this is consistent with the assumption that the bosonic Dirac operator $D_B$  made from $q_B$ is required to be self-adjoint, in the spectral action.
$p_B$ is 
\begin{equation}
    p_B = \frac{a}{2} (2\dot{q}_B +(\beta_1+\beta_2)\dot{q}_F)
\end{equation}
Assuming that $q_B$ is self adjoint, $p_B$ becomes self-adjoint when 
\begin{equation}
    [(\beta_1 \dot{q}_F)_{AS}+(\beta_2 \dot{q}_F)_{AS}] =0
\end{equation}
which also means
\begin{equation}
    (\beta_1+\beta_2)\dot{q}_F +\dot{q}_F^\dagger (\beta_1+\beta_2)^\dagger =0 \label{condition}
\end{equation}
Eqn. [\ref{condition}] does not say anything about the adjointness of $\beta_1$ and $\beta_2$ individually. We  may assume for simplicity that $\beta_1$ and $\beta_2$ are self-adjoint: 
\begin{equation}
    \beta_1^\dagger = \beta_1 \quad \& \quad \beta_2^\dagger = \beta_2
    \label{betacond}
\end{equation}
Using  [\ref{betacond}], [\ref{condition}] becomes
\begin{equation}
    \{(\beta_1+\beta_2),\dot{q}_{FS}\}+[(\beta_1+\beta_2),\dot{q}_{FAS}] = 0
\end{equation}
This is the condition for the adjointness of $p_B$. Next, self and anti-self adjoint parts of $p_F$ are given as 
\begin{align}
    p_{FS} &= \frac{a}{4} \Big[\{\dot{q}_B,(\beta_1+\beta_2) \} +2(\beta_1 \dot{q}_{FAS} \beta_2 +\beta_2 \dot{q}_{FAS} \beta_1) \Big] \\
    p_{FAS} &= \frac{a}{4} \Big[[\dot{q}_B,(\beta_1+\beta_2)]  +2(\beta_1 \dot{q}_{FS} \beta_2 +\beta_2 \dot{q}_{FS} \beta_1) \Big]
\end{align}
The self-adjoint and anti-self adjoint parts of the fermionic anti-commutator in the Adler-Millard charge are
\begin{align}
    \Tilde{C}_{FS} &= \{q_{S},(\beta_1\dot{q}_{AS} \beta_2 +\beta_2\dot{q}_{AS} \beta_1)\}+\{q_{AS},(\beta_1\dot{q}_{S} \beta_2 +\beta_2\dot{q}_{S} \beta_1)\} \\
    \Tilde{C}_{FAS} &= \{q_{S},(\beta_1\dot{q}_{S} \beta_2 +\beta_2\dot{q}_{S} \beta_1)\}+\{q_{AS},(\beta_1\dot{q}_{AS} \beta_2 +\beta_2\dot{q}_{AS} \beta_1)\}
\end{align}
In these two equations, $q$ stands for $q_F$. In [\ref{condition}], the sum of the terms is zero. However, if the terms are independently zero, i.e. $(\beta_1\dot{q}_F)_{AS}=0$ and $(\beta_2 \dot{q}_F)_{AS}=0$ then $\textbf{H}_{AS}$ will vanish.

In summary, we see that while it is reasonable to take $\dot{q}_B$ as self-adjoint, it is not really necessary to assume $p_B$, $p_F$ and $H$ to be self-adjoint, at Level 0. All that we should require is that at Levels II and III - i.e. in quantum field theory and classical dynamics, these quantities should become self-adjoint. We will see in the next section that this can be ensured. However, we do not make our measurements at Level 0; hence there is no reason to require these quantities to be self-adjoint at Level 0. In fact, it is their anti-self-adjoint component, which arises very naturally, that is responsible for a dynamical origin of the quantum-to-classical transition [spontaneous localisation]. Thus, in this theory, it is not necessary to ascribe any interpretation to quantum theory, to get the classical world to emerge from quantum theory. It is the presence of these anti-self-adjoint terms that gets missed when we construct a quantum theory of gravity by quantizing a classical theory of gravity. The bottom-up approach to quantum gravity has more information than a top-down approach.

The fact that the STM atom evolves like a free particle, inspite of the Hamiltonian not being self-adjoint, suggests that we could think of its motion in the Hilbert space as `geodesic' motion in a non-commutative geometry. We can associate a state vector with the STM atom, analogous to the four-velocity vector in special relativity, whose length remains unchanged during geodesic [i.e. free] motion.
This observation will help us in the next section to motivate the constancy of the norm of the state vector in the emergent theory. This overcomes a limitation of collapse models, in which norm preservation in the presence of stochastic noise has to be added by hand as an ad hoc assumption, so as to be able to derive the Born probability rule.

Our theory also enables us to construct a relativistic quantum (field) theory of spontaneous localisation. It is our contention that a relativistic theory of spontaneous collapse must treat time at the same footing as three-space. This implies that there must take place spontaneous localisation in coordinate time, besides in space. This requires us to treat coordinate time, besides the spatial position of a particle, as an operator. The role of time as an evolution parameter has to be then played by something else, and Connes time does precisely that. A relativistic quantum field theory must treat coordinate time also as an operator, but so long as spontaneous collapse in time can be neglected, treating operator time as a classical Lorentz invariant coordinate time is an excellent approximation, as is assumed in conventional quantum field theory. It is well-known though, that one can also develop an equivalent version of quantum field theory (the so-called Stueckelberg-Horwitz relativistic quantum mechanics \cite{stueck}) which treats time as an operator, and introduces alongwith, a new absolute time parameter for defining evolution.

At Level 0, the Hilbert space is populated by a large number of STM atoms, each of which is a free particle described by the dynamics described above. Interaction between atoms is via entanglement of their individual states. Each $q$-particle carries its own set of non-commuting space-time coordinates [Paper I]. There is no classical space-time; only a Hilbert space in which evolution is with respect to Connes time $\tau$. 
There is a conserved Adler-Millard charge for the collection of atoms, as given by (\ref{amc}), where the index $r$ indicates sum over all STM atoms.
Classical space-time emerges after one carries out a statistical thermodynamics of a large number of STM atoms, and spontaneous localisation arises away from thermodynamical equilibrium. This is described in the next section. Note that we do not quantise this matrix dynamics; rather, quantum theory emerges from it, just like in trace dynamics.

\section{The origin of spontaneous localisation}
Once the matrix dynamics at Level 0 has been specified by prescribing the Lagrangian, one constructs the statistical thermodynamics of a large number of STM atoms. The motivation is that if one is not observing the microscopic dynamics at the Planck scale, it is then the emergent coarse-grained dynamics which is of interest. To do this, one applies the standard principles of statistical mechanics to an ensemble of STM atoms, as is done in trace dynamics (see e.g. Chapter 4 of Adler's book \cite{Adler:04}). One starts by setting up an integration measure in the operator phase space for the bosonic and fermionic matrices. Then a Liouville theorem is derived. Next, given the operator phase space measure, one defines an equilibrium phase space density $\rho$, which is used to define the probability of finding the system in the phase space volume element $d\mu$. A canonical ensemble and an entropy function is constructed, as a function of the conserved charges: the trace Hamiltonian and the Adler-Millard charge. The equilibrium distribution is constructed by maximising the entropy function. While we will describe this analysis in detail in a forthcoming work, the analysis essentially follows that in trace dynamics. All that we have done in the present paper is to propose a specific trace dynamics Lagrangian which brings gravity into the trace dynamics framework, and unifies it with matter fermions. And although classical spacetime is lost at Level 0, Connes time enables us to define evolution.

This sets the stage for the emergence of the coarse-grained quantum gravitational dynamics at thermodynamic equilibrium. A Ward identity, which is the equivalent of the equipartition theorem, is derived. As in trace dynamics, the anti-self adjoint part of the conserved Adler-Millard charge is equipartitioned over all the degrees of freedom, and the equipartitioned value per degree of freedom is identified with Planck's constant $\hbar$. At thermodynamic equilibrium, the standard quantum commutation relations of (an equivalent of) quantum general relativity emerge, for the canonical averages of the various degrees of freedom:
\begin{equation}
[q_B, p_B] = i \hbar; \qquad \{q_{FS}, p^f_{FAS}\} = i\hbar; \qquad \{q_{FAS}, p^f_{FS}\} = i\hbar
\end{equation}
All the other commutators and anti-commutators amongst the canonical degrees of freedom vanish at thermodynamic equilibrium. The above set of commutation relations hold for every STM atom. We note that we describe quantum general relativity in terms of these $q$ operators, and not in terms of the metric and its conjugate momenta, which are emergent concepts of Levels II and III.

The mass $m$ of the STM atom is defined by $m=\hbar/Lc$; and $L$ is interpreted to be its Compton wavelength. Newton's gravitational constant $G$ is defined by $G\equiv L_p^2 c^3/\hbar$, and Planck mass $m_P$ by $m_P=\hbar/L_P c$. Mass and spin are both emergent concepts of Level I; at Level 0 the STM atom only has an associated length $L$.

As a consequence of Hamilton's equations at Level 0, and as a consequence of the Ward identity mentioned above, the canonical thermal averages of the canonical variables obey the Heisenberg equations of motion of quantum theory, these being determined by ${ H}_S$, the canonical average of the self-adjoint part of the Hamiltonian:
\begin{equation} 
i\hbar \frac{\partial q_B}{\partial \tau} = [q_B, { H}_S]; \qquad i\hbar \frac{\partial p_B}{\partial \tau} =  [p_B, { H}_S]; \qquad i\hbar \frac{\partial q_F}{\partial \tau} = [q_F,{ H}_S]; \qquad i\hbar \frac{\partial p^f_F}{\partial \tau} =  [p^f_F, { H}_S]
\end{equation}
In analogy with quantum field theory, one can transform from the above Heisenberg picture, and write a Schr\"{o}dinger equation for the wave-function $\Psi(\tau)$ of the full system:
\begin{equation}
i\hbar \; \frac{\partial \Psi}{\partial \tau} = { H_{Stot}} \Psi (\tau)
\end{equation}
where ${ H_{Stot}}$ is the sum of the self-adjoint parts of the Hamiltonians of the individual STM atoms. Since the Hamiltonian is self-adjoint the norm of the state vector is preserved during evolution.  This equation is the analog of the Wheeler-DeWitt equation in our theory, the equation being valid at thermodynamic equilibrium at Level I. This equation can possibly resolve the problem of time in quantum general relativity, because to our understanding it does not seem necessary that the physical state must be annihilated by ${ H_{Stot}}$. We have not arrived at this theory by quantising classical general relativity; rather the classical theory will emerge from here after spontaneous localisation, as we now describe.

We can now describe  how spontaneous localisation comes about. It is known that the above emergence of quantum dynamics arises at equilibrium in the approximation that the Adler-Millard conserved charge is anti-self-adjoint, and its sef-adjoint part can be neglected. In this  approximation, the Hamiltonian is self-adjoint. Another way of saying this is that quantum dynamics arises when statistical fluctuations around equilibrium (which are governed by the self-adjoint part of $\tilde{C}$) can be neglected. When the thermodynamical fluctuations are important, one must represent them by adding a stochastic anti-self-adjoint operator function to the total self-adjoint Hamiltonian (note that one cannot simply add the anti-self-adjoint part of the Hamiltonian to the above Schr\"{o}dinger equation, because that equation is defined for canonically averaged quantities; the only way to bring in fluctuations about equilibrium is to represent them by stochastic functions). This way of motivating spontaneous collapse is just as in trace dynamics (see Chapter 6 of \cite{Adler:04}), except that we are not restricted to the non-relativistic case, and evolution is with respect to Connes time $\tau$. Also, we do not have a classical space-time background yet; this will emerge now, as a consequence of spontaneous localisation [see also our earlier related paper `{\it Space-time from collapse of the wave function}' \cite{Singh:2019}].

Thus we can represent the inclusion of the anti-self-adjoint fluctuations in the above Schr\"{o}dinger equation by a stochastic function ${\cal H}(\tau)$ as:
\begin{equation}
i\hbar \; \frac{\partial \Psi}{\partial \tau} = [{ H_{Stot}} + {\cal H}(\tau)] \Psi (\tau)
\end{equation}
In general, this equation will not preserve norm of the state vector during evolution. However, as we noted above, every STM atom is in free particle geodesic motion. Hence it is very reasonable to demand that the state vector should preserve norm during evolution, even after the stochastic fluctuations have been added. Then, exactly as in collapse models and in trace dynamics, a new state vector is defined, by dividing $\Psi$ by its norm, so that the new state vector preserves norm. Then it follows that the new norm preserving state vector obeys an equation which gives rise to spontaneous localisation, just as in trace dynamics and collapse models (see Chapter 6 of \cite{Adler:04}). We should also mention that the gravitational origin of the anti-self-adjoint fluctuations presented here ($D_F$ is likely of gravitational origin, and relates to the anti-symmetric part of an asymmetric metric) agrees with Adler's proposal that the stochastic noise in collapse models is seeded by an imaginary component of the metric \cite{Adler2014, Adler:2018}.

It turns out to be rewarding to work in the momentum basis where the state vector is labelled by the eigenvalues of the momenta $p_B$ and $p_F$. Since the Hamiltonian depends only on the momenta, the anti-self adjoint fluctuation is determined by the anti-self adjoint part of $p_F$. Hence it is reasonable to assume that spontaneous localisation takes place onto one or the other eigenvalue of $p^f_F$. No localisation takes place in $p_B$ - this helps understand the long range nature of gravity (which results from $q_B$ and the bosonic Dirac operator $D_B$). We assume that the localisation of $p^f_F$ is accompanied by the localisation of $q_F$, and hence that an emergent classical space-time 
is defined using the eigenvalues of $q_F$ as reference points.  Space-time emerges only as a consequence of the spontaneous localisation of matter fermions. Thus we are proposing that the eigenvalues of $q_F$ serve to define the space-time manifold. It is not clear to us at this stage as to what exactly is the relation between the $q$-operator and the classical space-time metric: as of now we assume that when spontaneous localisation leads to the emergence of a classical space-time, it also (somehow) defines the space-time metric.
 As in collapse models, the rate of localisation becomes significant only for objects which comprise a large number of matter fermions - hence the emergence of a classical space-time is possible only when a sufficiently macroscopic object comprising many STM atoms undergoes spontaneous localisation. It is evident that such localisation is a far from equilibrium process, consequent upon a sufficiently large statistical fluctuation coming into play. We now give a quantitative estimate as to what qualifies as sufficiently macroscopic. 

To arrive at these estimates, we recall the following two earlier equations, the action principle for the STM atom itself, and the eigenvalue equation for the full Dirac operator $D$:
\begin{equation}
\frac{L_P}{c} \frac{S}{\hbar}  =  \frac{a}{2} \int d\tau \; Tr \bigg[ \dot{q}_B^2 +\dot{q}_B\beta_2\dot{q}_F +\beta_1 \dot{q}_F \dot{q}_B +\beta_1 \dot{q}_F \beta_2 \dot{q}_F \bigg]
\label{newacnr}
\end{equation}
\begin{equation}
[D_B + D_F] \psi = \lambda \psi \equiv (\lambda_R + i \lambda_I)\psi \equiv \bigg(\frac{1}{L} + i \frac{1}{L_I}\bigg)\psi
\end{equation}
In the second equation. since $D$ is bosonic, we have assumed that the eigenvalues $\lambda$ are complex numbers, and separated each eigenvalue into its real and imaginary part. Furthermore, this will be taken as the definition of the length scale $L$ introduced earlier. We come back to $L_I$ below. There will be one such pair of equations for each STM atom, and the total action of all STM atoms will be the sum of their individual actions, with the individual action given as above.

When an STM atom undergoes spontaneous localisation, $p^f_F$ localises to a specific eigenvalue. Since $D_F$ is also made from $\dot{q}_F$, just as $p^f_F$ is, we assume that $D_F$ also localises to a specific eigenvalue, whose imaginary part is the $L_I$ introduced above. Correspondingly, the $D_B$ associated with this STM atom acquires a real eigenvalue, which we identify with the $\lambda_R\equiv 1/L$ above (setting aside for the moment the otherwise plausible situation that in general $p_F$ will also contribute to $\lambda_R$). 

The spontaneous localisation of each STM atom to a specific eigenvalue reduces the first term of the trace Lagrangian to:
\begin{equation}
Tr [\dot{q}_B^2] \rightarrow \lambda_R^2
\end{equation}
If sufficiently many STM atoms undergo spontaneous localisation to occupy the various eigenvalues $\lambda^i_R$  of the Dirac operator $D_B$, then we can conclude, from our knowledge of the spectral action in non-commutative geometry \cite{Landi1999}, that their net contribution to the trace is:
\begin{equation} 
\frac{\hbar a}{2} \;  Tr[\dot{q}_{B}^2] =\frac{\hbar }{2}  Tr [L_p^2 D_B^2] = \frac{\hbar}{2} L_p^2 \sum  (\lambda_R^i)^2 = \frac{\hbar}{2 L_p^2} \int d^4x \; \sqrt{g} \; R
\end{equation}
Thus we conclude that the Einstein-Hilbert action emerges after spontaneous localisation of the matter fermions. In that sense, gravitation is indeed an emergent phenomenon. Also, the eigenvalues of the Dirac operator $D_B$ have been proposed as dynamical observables for general relativity \cite{Rovelli}, which in our opinion is a result of great significance.

Let us now examine how the matter part of the general relativity action arises from the trace Lagrangian (its second and third terms) arises after spontaneous localisation. These terms are
\begin{equation}
\frac{a\hbar }{2}  Tr \big[ \dot{q}_B\beta_2\dot{q}_F +\beta_1 \dot{q}_F \dot{q}_B \big] = \hbar Tr [L_p^2 D_F D_B]
\end{equation}
Spontaneous localisation sends this term to $L_p^2 \times {1}/{L_I} \times {1}/{L}$. There will be one such term for each STM atom, and analogous to the case of $Tr D_B^2$ we anticipate that the trace over all STM atoms gives rise to the `source term'
\begin{equation}
\hbar\int \sqrt{g} \; d^4x \;\sum_i [ L_p^{-2} \times {1}/{L^i_I} \times {1}/{L^i}]
\end{equation}
Consider the term for one atom. We make the assumption (which becomes plausible shortly) that spontaneous localisation localises the STM atom to a spatial volume $L^3$ such that $L_p^2 L_I = L^3$. We note that it is natural to identify $L$ with the Compton wavelength $\hbar/ mc$ of the STM atom. Moreover, we may say that the classical approximation consists of replacing the inverse of the spatial volume of the localised particle - $1/L^3$, by the spatial delta function $\delta^3({\bf x} - {\bf x_0})$ so that the contribution to the matter source action becomes
\begin{equation}
\hbar \int \sqrt{g} \; d^4x \; [ L_p^{-2} \times {1}/{L_I} \times {1}/{L}] = mc \int ds
\end{equation}
which of course is the action for a relativistic point particle.

Putting everything together, we conclude that upon spontaneous localisation, the fundamental trace based action for a collection of STM atoms becomes
\begin{equation}
S = \int d^4x\; \sqrt{g} \; \bigg [\frac{c^3}{2G}R + c\; \sum_i m_i \delta^3({\bf x} - {\bf x_0})\bigg]
\end{equation}
In this way, we recover general relativity at Level III, as a result of spontaneous localisation of quantum general relativity at Level I. We should not think of the gravitational field of the STM atom as being disjoint from its related fermionic source: they both come from the same eigenvalue $\lambda$, being respectively the real and imaginary parts of this eigenvalue.

Strictly speaking, the Connes time integral should also be displayed in the action principle:
\begin{equation}
S = \frac{c}{L_P} \int d\tau\; \int d^4x\; \sqrt{g} \; \bigg [\frac{c^3}{2G}R + c\; \sum_i m_i \delta^3({\bf x} - {\bf x_0})\bigg]
\end{equation}
It is as if the observed universe is an enormous spontaneously collapsed bubble that evolves `inside of a sea of' uncollapsed STM atoms. Inside the bubble there is a spacetime, with its own time evolution parameter, with no direct indicator of Connes time. `Outside' of the bubble there is no space-time, but only a Hilbert space populated with other STM atoms, evolving in Connes time. Could it be that the big bang represents an exceedingly  huge spontaneous collapse event, involving an entangled state of an astronomical number of STM atoms? Is such spontaneous localisation accompanied by the expansion of the resulting classical space-time? And could it be that there are very many other spontaneously collapsing bubble universes forming all the (Connes) time, in the Hilbert space of STM atoms? The far from equilibrium dynamics of such spontaneous fluctuations in an ensemble of STM atoms should be an interesting aspect to explore.

We have not been able to come to a definite conclusion as regards what happens to the last term in the trace Lagrangian (\ref{newacnr}), (i.e.  $\beta_1 \dot{q}_F \beta_2 \dot{q}_F$) - after spontaneous localisation. It roughly has the structure $Tr [D_F^2]$. Adding the contribution of the eigenvalues $q_{F1}$ and $q_{F2}$ of $\beta_1 \dot{q}_F$ and $\beta_2 \dot{q}_F$ from all STM atoms, we get $Tr [q_{F1i} \; q_{F2i}]$. While we do not have a proof, we suggest that this could give rise to the cosmological constant term of general relativity. If this were to be true, then we can schematically sum up the overall picture as
\begin{equation}
S_{NMG} = \int d\tau \sum_i Tr D^2_i  \quad  {\mathbf \longrightarrow} \quad \int d\tau \bigg[ \frac{c^3}{2G}\int d^4x \; \sqrt{g} \; [R-2\Lambda] + \int d^4x \; \sqrt{g} \; L_{matter}\bigg]
\end{equation}
Here, $S_{NMG}$ on the left is the total action of all STM atoms in this Non-commutative Matter-Gravity. The action on the right side of the arrow describes classical general relativity with a cosmological constant and point matter sources, and is what emerges after spontaneous localisation.  Our theory thus elegantly unifies in a simple way, the disjoint matter - gravity descriptions on the right hand side, by bringing them together as $\sum Tr\; D_i^2$. Note that, unlike the action on the left hand side of the arrow,  the right hand side of the above equation is in no way the sum of contribution of individual STM atoms: the matter part is a sum, but the gravity part is not. Undoubtedly then, the gravity part is an emergent condensate. It simply cannot be quantised. The right hand is the [commutative] action at Level III covariant under general coordinate transformation of commuting coordinates. Whereas the left hand side action at Level I is covariant under general coordinate transformations of {\it non-commuting} coordinates. It is interesting that the transition from a non-commutative geometry to a commutative geometry is caused by spontaneous localisation, and that statistical thermodynamics plays a central role in it.

Let us return to our assumption $L_I = L^3/L_P^2$, which as we will now see, has profound consequences. If we consider the case of a nucleon, and substitute for $L$ the Compton wavelength of a nucleon ($\sim 10^{-13}$ cm), then $L_I$ comes out to be $10^{27}$ cm, which is close to the size of the observed universe ($10^{28}$ cm). If this is not a coincidence, it suggests the possibility of a connection between spontaneous localisation and the scale of the universe. In particular, it might be possible to define the rate of spontaneous collapse as $L_I/c\sim 10^{-17}$ sec$^{-1}$, which happens to be the same as the collapse rate assumed in standard models of localisation.  

Let us return next to the modified Dirac equation that we wrote above:
\begin{equation}
[D_B + D_F] \psi = \lambda \psi \equiv (\lambda_R + i \lambda_I)\psi \equiv \bigg(\frac{1}{L} + i \frac{1}{L_I}\bigg)\psi
\end{equation}
Substituting for $L_I$ as $L_I = L^3 / L_P^2$ we can write this as
\begin{equation}
[D_B + D_F] \psi =  \frac{1}{L} \bigg(1+ i \frac{L_P^2}{L^2}\bigg)\psi
\label{mod}
\end{equation}
and it is instructive to define a complex length scale $L_{com}$ by
\begin{equation}
\frac{L_{com}}{L^2{}} = \frac{1}{L} \bigg(1+ i \frac{L_P^2}{L^2}\bigg) \equiv \frac{1}{L} \bigg(1+ i \frac{R_S}{L}\bigg) \implies L_{com} = L + i R_S
\end{equation}
where $R_S$ is the Schwarzschild radius associated with an STM atom of mass $m$ having a Compton wavelength $1/L$. If $R_S \ll L$, (i.e. $m\ll m_{Pl}$), the imaginary part of the complex length is ignorable compared to the real part, hence $D_F$ can be ignored compared to $D_B$ and we get the standard Dirac equation for a matter fermion. This is the microscopic quantum limit, where spontaneous localisation is insignificant. On the other hand, when $R_S \gg L$, (i.e. $m\gg m_{Pl}$), the imaginary part of the length $L_{com}$ dominates over the real part, spontaneous localisation is significant, and we recover classical behaviour: in fact, a black hole solution of radius $R_S$.  Thus our matrix dynamics nicely interpolates between quantum theory and classical mechanics; we did not have to put in this behaviour by hand, rather it comes out quite naturally from the theory. Incidentally, the ratio $L^2/R_S=\hbar^2/Gm^3$ which is implicit in the above length scale, arises naturally as the decoherence length scale when one studies gravitationally induced decoherence using the Schr\"{o}dinger-Newton equation. In our context, it implies that gravitational decoherence is significant when Compton wavelength is larger than this decoherence length. As expected, for a nucleon this length is of the order of the size of the universe. The length $L_{com}$ can not have a magnitude smaller than Planck length - this possibly is a way of avoiding the gravitational singularity of a black hole, because it will never shrink to a vanishing $L_{com}$. 

For a black hole made of $N$ nucleons, the amplified collapse rate is $10^{-17}\; N$ sec$^{-1}$, which for an astrophysical black hole with $N\sim 10^{57}$ gives an extremely rapid rate of $10^{40}$ collapses per second - obviously then, black holes behave classically. An object with a mass smaller than Planck mass cannot be a black hole: it is necessarily quantum in nature.

The Dirac equation (\ref{mod}) is one of great significance, as it admits solutions which have Dirac fermions and black holes as their special limiting cases. In that sense, this equation unifies the standard Dirac equation with Einstein equations. It answers the question: given a relativistic mass $m$, does it obey the Dirac equation or Einstein's equations? It also helps understand why a Kerr-Newman black hole has the same gyromagnetic ratio as the electron: because they are both solutions of the same equation - i.e. Eqn. (\ref{mod}). 

Equation (\ref{mod}) admits a duality, between black hole and fermion solutions, after it is written as
\begin{equation}
[D_B + D_F] \psi =  \frac{L_{com}}{L^2} \psi
\label{mod2}
\end{equation}
Given a black hole solution $\psi_{BH}$, (i.e. $R_S \gg L$) with mass $m_{BH}$, consider a Dirac fermion solution $\psi_{F}$ (i.e. $L'\gg R'_S$) with a mass $m_F = m^2_{Pl}/m_{BH}$. That is, the second solution is obtained by setting its Schwarzschild radius equal to Compton wavelength of first solution, and setting its Compton wavelength equal to Schwarzschild radius of first one. This interchange means that the eigenvalue $\lambda_F$ of the second solution is $m_{Pl}^2/m^2_{BH}$ times the eigenvalue $\lambda_{BH}$ of the first one, and the real and imaginary parts have been interchanged. Thus, given a solution $\psi_{BH}$ with eigenvalue $\lambda_{BH}$, its dual solution
$\psi_F$ can be found by interchanging the real and imaginary parts of the eigenvalue, and by downscaling the magnitude of the eigenvalue as just mentioned. This duality might be of some help in a computation of the entropy of the black hole, from a knowledge of the eigenvalues and eigenstates of the Dirac operator $D_B$. This is plausible because the action function of the black hole, which in turn is related to its entropy, is after all constructed from the eigenvalues of $D_B$.

The duality $L\rightarrow L_P^2/L$ has been investigated earlier as well. The conclusion that space-time intervals have a minimum length $L_P$ has been derived  in \cite{paddy97}  by proposing that in the path integral for a spin-zero free particle, paths of length $l$ have the same weightage as paths of length $L_P^2/l$. 
The subtle difference in our case is the $i$ factor, which enables spontaneous localisation: thus the duality in our case is $l\rightarrow i L_P^2/l$.

Thus far we have derived a reasonable understanding of the dynamics of STM atoms at Level 0, Level I and Level III. Level 0 is the fundamental matrix dynamics; Level I is its statistical thermodynamics (equivalent of quantum general relativity) and Level III is its classical limit [caused by spontaneous localisation] - i.e. the classical theory of general relativity. Level II is the hybrid level: quantum field theory on a curved space-time. Level II is concerned with those STM atoms which have not yet undergone spontaneous localisation - how to describe their dynamics from the point of view of the classical space-time which has already been created from the spontaneous localisation of many many other STM atoms? These uncollapsed atoms obviously live at Level 0, where we know how to understand their dynamics. Furthermore, we can also do the statistical thermodynamics for them and describe them at Level I, while neglecting their spontaneous localisation at Level I. To arrive at Level II, we first neglect their own gravitational degree of freedom $q_B$; replace the Connes time evolution by time evolution provided by the emergent space-time of collapsed STM atoms, and replace their $D_B$ operator by the standard Dirac operator associated with the emergent space-time. We also neglect $D_F$ (no spontaneous localisation). Thus we have the standard Dirac equation for fermions, as well their standard quantum commutators. Since we have not yet introduced non-gravitational interactions in this matrix dynamics, there are no other bosonic fields yet in the theory.

It is very interesting to ask if we could have missed out some vital information in going from Level I to Level II in the above manner. Indeed we have. We recall from the discussion earlier in this section that space-time is emergent from the spontaneous localisation of $q_F$. If we were to describe spontaneous localisation of the fermionic degrees of freedom at Level II, we must invoke the statistical fluctuations around equilibrium. And if we want to have a relativistic theory of collapse at Level II, we will need to bring Connes time back into the picture, and describe spontaneous localisation at Level II precisely as we did above. We cannot appeal to the emergent space-time to provide a background for relativistic collapse \cite{Singh:2019}. This leads to a falsifiable prediction: spontaneous collapse of the operator coordinate time $\hat{t}$. Thus Connes time is crucial at Level II as well, if we are to describe relativistic spontaneous localisation. Of course one can take the non-relativistic limit of the relativistic theory of Level II, and then describe collapse in absolute Newtonian time, as is done in conventional collapse models.

\section{Concluding Remarks}
We have presented a viable new quantum theory of gravity, which predicts spontaneous localisation, and spontaneous collapse in time. The theory is hence falsifiable, and a vigorous experimental effort is currently underway in various laboratories to test models of spontaneous collapse \cite{RMP:2012, Mauro2019}. Our theory combines non-commutative geometry and trace dynamics to construct a matrix dynamics for our newly introduced concept of atoms of space-time matter. This is a quantum theory of gravity, from which quantum general relativity and classical general relativity are emergent as  approximations.

Amongst outstanding open issues which still need to be resolved is to understand the exact relation between the $q$-operator and the space-time metric, and the interpretation of the operator $D_F$. Our guess is that $q_B$ somehow relates to the symmetric part of an asymmetric metric at Level 0, and $q_F$ relates to its complex anti-symmetric part; with $D_F$ being related to a complex torsion induced by the anti-symmetric part of the asymmetric metric. Also, right at the beginning we made the assumption $\chi(u)=u$ and restricted ourselves to the second order in the heat kernel expansion of the Dirac operator. Also we neglected the cosmological constant term which arises at order $L_P^{-4}$ in the heat kernel expansion.These assumptions will need to be relaxed, so that one deals with the full theory without an expansion in $L_P^2$. Furthermore, $D_B$ is defined on a Euclidean space-time, so we have a Euclidean quantum gravity theory. This will have to be replaced by the Lorentzian theory. Another important aspect is to now include other interactions in this framework.

It would be of great interest to explore the eigenvalues and eigenstates of the full Dirac operator $D=D_B+D_F$ on a non-commutative space. This could help predict the discrete masses of elementary particles. We have already seen above the relation $L_I = L^3/L_P^2$ between the real and imaginary parts of the eigenvalue of the $D$-operator. The fact that $L_I$ comes out to be the order of the size of the observed universe, introduces an infra-red scale into the theory, which could help address the hierarchy problem. In fact this relation allows us to `predict' the mass $m_{Pr}$ of a proton in terms of Planck mass $m_P$ and the Hubble parameter ($L_I\sim cH_{0}^{-1}$), recalling that $L$ is the Compton wavelength:
\begin{equation}
\frac {m_{Pr}}{m_P} \approx \bigg( \frac{L_P}{cH_{0}^{-1}}\bigg)^{1/3}
\end{equation}
We can say that the proton mass is much much smaller than Planck scale, because the universe is much much bigger than the Planck scale.

Our theory probably also has interesting implications for black-hole evaporation. Black holes for us arise as a spontaneous localisation of a collection of STM atoms having a total mass $M$, for which the Schwarzschild radius exceeds the Compton wavelength. Thus black hole formation is a far from equilibrium non-unitary (caused by a statistical fluctuation) process. Thus even though a black hole has enormous entropy, it is nonetheless a far from-equilibrium state of relatively low entropy (compared to what it would be at thermodynamic equilibrium). Hawking evaporation is a process (opposite to spontaneous localisation) whereby a black hole returns to thermodynamic equilibrium with the `sea of STM atoms' in the Hilbert space. Recall that the equilibrium is described by quantum general relativity / quantum field theory. In the long run, all matter in the universe will condense into black holes, and then evaporate as radiation and go back to quantum gravitational equilibrium - this will amount to loss of classical spacetime, and a return to evolution in Connes time. There is no question of an information loss paradox, because in the first place the formation of a black hole is itself a non-unitary process
 \cite{Singh:2019essay}.

Lastly we note that we were compelled to introduce two constant matrices $\beta_1$ and $\beta_2$, and work with $q_B + \beta_1 q_F$ and $q_B + \beta_2 q_F$. As if to suggest that our STM atom is a two dimensional entity [`space time is two-dimensional at the Planck scale?']. Could it be that the object that we have called the STM atom is after all the closed string of classical string theory? Or could it be related to the loop in loop quantum gravity?

\newpage

\noindent {\bf Acknowledgements:} 
For valuable discussions and ongoing collaboration we would like to thank 
B. J. Arav, 
Shivnag Sista, 
Diya Kamnani,
Guru Kalyan, 
Shubham Kadian, 
Siddhant Kumar, 
Ankur, 
Abhinav Varma,
Ish Mohan Gupta,
Krishnanand K Nair and
Shounak De. TPS would like to thank Thanu Padmanabhan for insightful conversations on this work.


\bigskip

\bigskip

\centerline{\bf REFERENCES}

\bibliography{biblioqmtstorsion}

\end{document}